\documentclass[a4paper, superscriptaddress, twocolumn, prl, showpacs, longbibliography]{revtex4-1}

\usepackage{dcolumn}
\usepackage{bm}
\usepackage[colorlinks=true,citecolor=blue,linkcolor=blue,pdfhighlight =/O]{hyperref}
\usepackage{graphicx}
\usepackage{amsmath}
\usepackage{cancel}

\usepackage{dsfont}
\usepackage{mathtools}
\usepackage[dvipsnames]{xcolor}
\usepackage{soul}
\usepackage[normalem]{ulem}
\usepackage{bm}
\usepackage{amsmath,amssymb,amsfonts,latexsym,fancyhdr,graphicx}
\usepackage{simplewick}
\usepackage[english]{babel}
\usepackage{graphicx}
\usepackage{subfigure}
\usepackage{xcolor}
\usepackage{color}
\usepackage{verbatim}

\newcommand{\elke}[1]{\textcolor{Purple}{#1}}
\hypersetup{urlcolor=blue}
\begin{document}

\title{Calorimetry of a Quantum Phase Slip }

\author{E. G\"um\"u\c{s}}
\affiliation{\mbox{Univ.} Grenoble Alpes, CNRS, Grenoble INP, Institut N\' eel, 25 rue des Martyrs, Grenoble, France}

\author{D. Majidi}
\affiliation{\mbox{Univ.} Grenoble Alpes, CNRS, Grenoble INP, Institut N\' eel, 25 rue des Martyrs, Grenoble, France}

\author{D. Nikoli{\'c}}
\affiliation{Fachbereich Physik, Universit\"at Konstanz, D-78457 Konstanz, Germany}

\author{P. Raif}
\affiliation{\mbox{Univ.} Grenoble Alpes, CNRS, Grenoble INP, Institut N\' eel, 25 rue des Martyrs, Grenoble, France}
\affiliation{Fachbereich Physik, Universit\"at Konstanz, D-78457 Konstanz, Germany}

\author{B. Karimi}
\affiliation{QTF Centre of Excellence, Department of Applied Physics, Aalto University School of Science, P.O. Box 13500, 00076 Aalto, Finland}

\author{J. T. Peltonen}
\affiliation{QTF Centre of Excellence, Department of Applied Physics, Aalto University School of Science, P.O. Box 13500, 00076 Aalto, Finland}

\author{E. Scheer}
\affiliation{Fachbereich Physik, Universit\"at Konstanz, D-78457 Konstanz, Germany}

\author{J. P. Pekola}
\affiliation{QTF Centre of Excellence, Department of Applied Physics, Aalto University School of Science, P.O. Box 13500, 00076 Aalto, Finland}

\author{H. Courtois}
\affiliation{\mbox{Univ.} Grenoble Alpes, CNRS, Grenoble INP, Institut N\' eel, 25 rue des Martyrs, Grenoble, France}

\author{W. Belzig}
\affiliation{Fachbereich Physik, Universit\"at Konstanz, D-78457 Konstanz, Germany}

\author{C. B. Winkelmann}
\affiliation{\mbox{Univ.} Grenoble Alpes, CNRS, Grenoble INP, Institut N\' eel, 25 rue des Martyrs, Grenoble, France}

\begin{abstract}
   In a Josephson junction, which is the central element in superconducting quantum technology, irreversibility arises from abrupt slips of the gauge-invariant quantum phase difference across the contact. A quantum phase slip (QPS) is often visualized as the tunneling of a flux quantum in the transverse direction to the superconducting weak link, which produces dissipation. 
In this work, we detect the instantaneous heat release caused by a QPS in a Josephson junction using time-resolved electron thermometry on a nanocalorimeter, signaled by an abrupt increase of the local electronic temperature in the weak link and subsequent relaxation back to equilibrium. Beyond providing a cornerstone in experimental quantum thermodynamics in form of observation of heat in an elementary quantum process, this result sets the ground for experimentally addressing the ubiquity of dissipation, including that in superconducting quantum sensors and qubits.
\end{abstract}

%\keywords{Suggested keywords}%Use showkeys class option if keyword

\selectlanguage{english}                             %display desired
\maketitle

The magnetic flux threading a superconducting loop is quantized in units of the flux quantum $\Phi_0=h/2e$. The tunneling of a flux quantum in or out of such a loop is associated to a change of $2\pi$ in the winding of the phase of the quantum wave function along the loop. The manipulation of individual flux quanta is at the core of superconducting circuit logics, both in the classical and in the quantum information regime \cite{mooij2005,mooij2006,liebermann2016}. Rapid single-flux-quantum (RSFQ) logic can operate up to 100 GHz frequencies and is considered as promising classical control electronics of qubits \cite{leonard_digital_2019,howington_interfacing_2019,mcdermott_quantumclassical_2018}.
In the quantum regime, the coherent superposition and manipulation of flux states is at the basis of flux qubits \cite{chiorescu2003,yan2016} and the fluxonium \cite{Manucharyan2009,nguyen2019}.

\begin{figure}
    \centering
    \includegraphics[width=1.1\columnwidth]{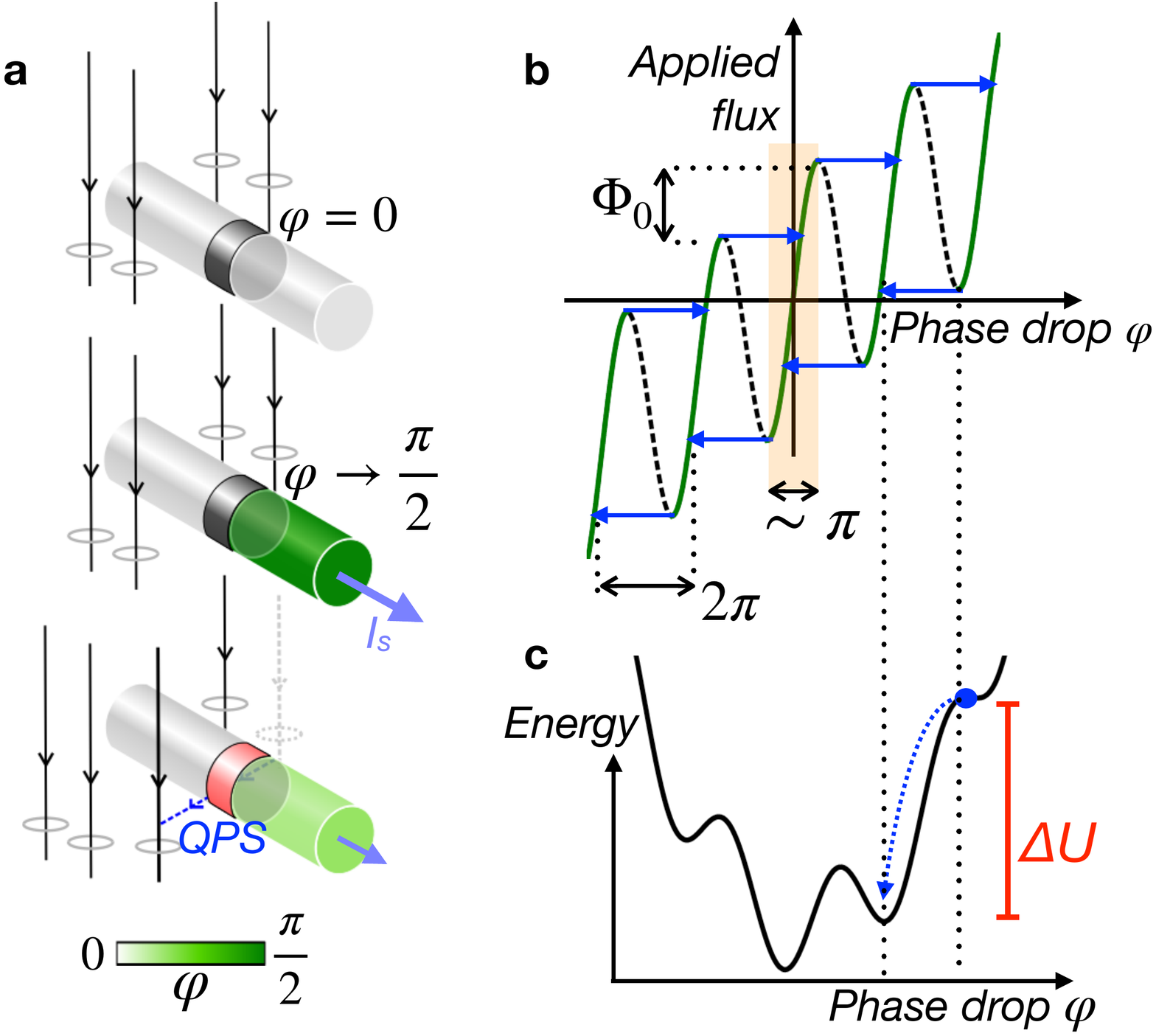}
     % \vspace{-1cm}
    \caption{{Quantum phase slip in a Josephson junction.} (a) Real-space sketch of the QPS mechanism: at the instability point of the $\Phi_x(\varphi)$ relation, the phase drop $\varphi$ and the screening current $I_s$ relax abruptly to smaller values, as a quantum of flux tunnels perpendicular to the Josephson junction (dark grey), releasing heat. (b) Phase drop $\varphi$ across the SNS junction versus applied flux to the SQUIPT, following \mbox{Eq.} (\ref{eq:phi-flux}) with $\beta=10$. The dashed part of the curve cannot be accessed. In a quantum phase slip (blue arrows), $\varphi$ changes by slightly less than $2\pi$. (c) Potential energy of the SQUIPT as a function of $\varphi$. A local energy minimum can become unstable as the externally applied flux is changed. By macroscopic quantum tunneling of the phase, a lower energy valley is reached, releasing an energy $\Delta U$. }
    \label{fig:schematics}
\end{figure}

Slips of the quantum phase occur when the gauge-invariant phase difference across a weak link, that is, a Josephson junction, in the superconducting loop suddenly relaxes (\mbox{Fig.} \ref{fig:schematics}). Quantum phase slips (QPSs) are ubiquitous in superconducting electronics and can be seen as the dual process to Cooper pair tunneling. Furthermore, coherent QPSs have been proposed as a building block for phase-slip qubit devices \cite{mooij2005,mooij2006,astafiev2012}.
On the other hand, their proliferation is responsible for the destruction of superconductivity in one dimension \cite{golubev2001} and can lead to thermal avalanches in current-biased Josephson junctions \cite{sahu2009}. In essence, a QPS can be considered the {\it quantum of dissipation} in superconducting electronics.

In this work, we investigate the thermal signature of incoherent individual quantum phase slips in a superconducting Josephson junction. Applying a MHz-bandwidth electron thermometry technique to a Josephson junction, we measure the heat generated by a single QPS as well as the subsequent thermal relaxation. The data are in good agreement with a theoretical model that we developed for describing the superconducting properties of the device. Our work therefore demonstrates the possibility to quantitatively account for dissipative effects in quantum nanoelectronics, with evident applications to quantum computing.
%\bigskip

The experimental core element is a superconducting quantum interference device (SQUID), that is, a superconducting loop containing one (as is the case here) or more Josephson junctions, and to which a magnetic flux $\Phi_x$ is applied. 
The difference between the applied and the physical flux is absorbed by screening supercurrents, leading to a gradient in the phase of the quantum wave function along the loop, and, to a large extent, across the Josephson weak link. 
The relation between $\Phi_x$ and the phase drop $\varphi$ across the Josephson junction can be written as
\begin{equation}
2\pi(\Phi_x/\Phi_0)= \varphi +\beta\sin(\varphi),
\label{eq:phi-flux}
\end{equation}
with the screening parameter $\beta=2\pi LI_c/\Phi_0$, where $L$ is the loop inductance and $I_c$ the Josephson junction's critical current \cite{tinkham1996dover}. 
Irreversibility arises when the SQUID's magnetic screening parameter $\beta$ exceeds 1 and \mbox{Eq.} (\ref{eq:phi-flux}) is no longer single-valued, as illustrated in \mbox{Fig.} \ref{fig:schematics}b,c.
In this situation, the penetration of an additional flux quantum into the SQUID loop does not occur smoothly and reversibly, but via a sudden tunneling of the phase, that is, a quantum phase slip. 
 
\begin{figure*}
    \centering
    %\vspace{-2cm}
    \includegraphics[width=0.98\textwidth]{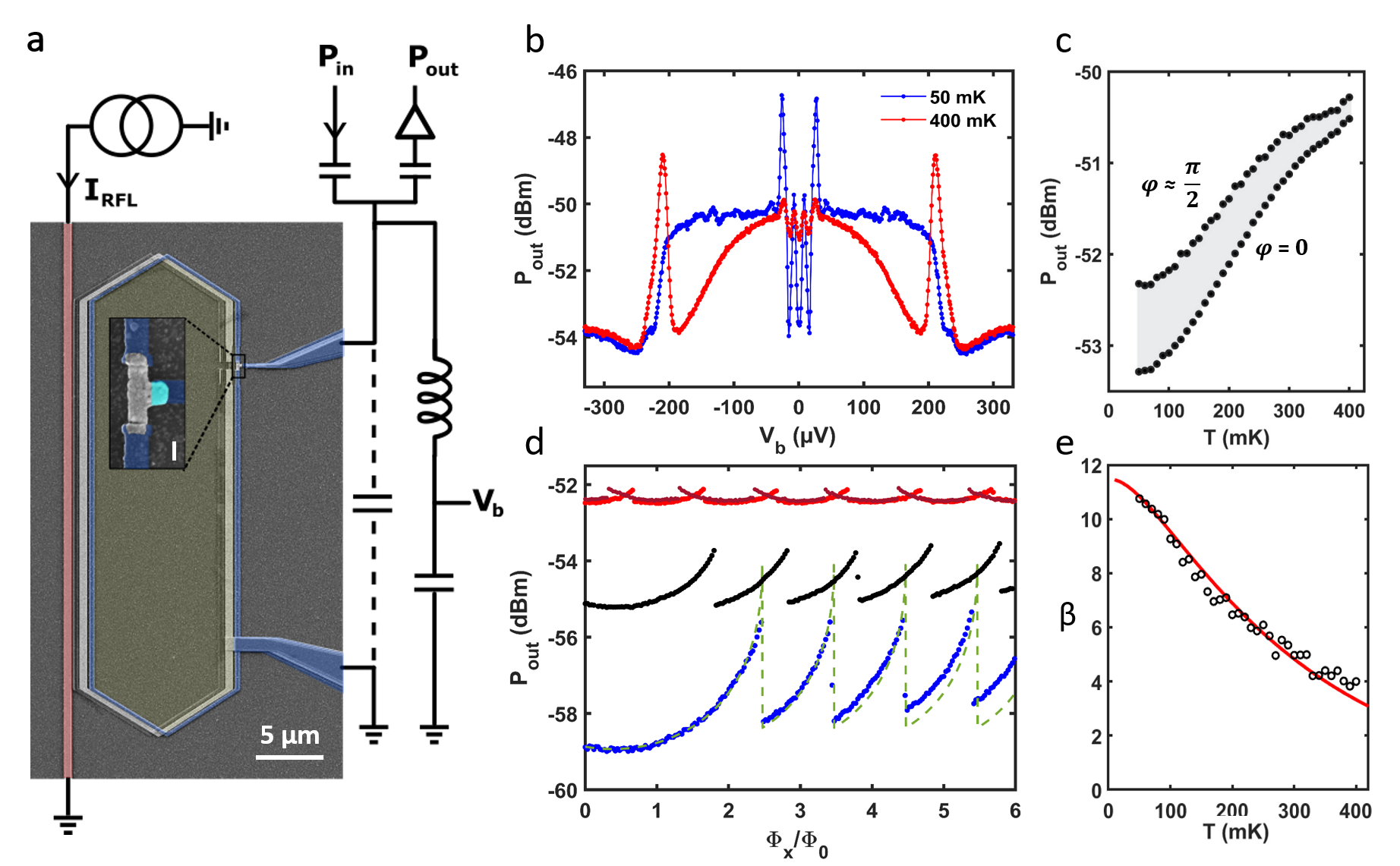}
    %SQUIPT-static
     % \vspace{-1cm}
    \caption{{Hysteretic RF-SQUIPT.} (a) Device schematics, including a false-color scanning electron micrograph (scale bar 5 $\mu$m) of the SQUIPT loop (yellow area) and the rapid flux line (pink). The zoom highlights the SNS junction (Al = blue, Cu = grey), connected laterally by the tunnel contact (cyan) (scale bar 200 nm), connected to the resonator. (b) Bias spectroscopy $P_{\rm out}(V_b)$ at cryostat temperatures 50 mK (blue) and 400 mK (red). (c) Calibration of $P_{\rm out}(V_b=0)$ versus cryostat temperature at equilibrium, under two different phase drops. The grey shaded region thus covers all possible values of $P_{\rm out}(T,\varphi)$. (d) Resonator response at $V_b=0$ as a function of increasing applied magnetic flux, at three cryostat temperatures (50 mK, blue; 200 mK, black; 400 mK, red). The dark red line exemplifies the response to a downward sweep of the flux. The dashed line is a calculation (see text). (e) Temperature dependence of the screening parameter $\beta$, extracted from (d), and theoretical fit (red line, see text).}
    \label{fig:static}
\end{figure*}

The above concepts are the basis of extensive applications in superconducting electronics, such as SQUID magnetometry and superconducting flux qubits \cite{vanderwal2000,friedman2000,chiorescu2003}.
Notably, the Josephson weak links used in such SQUIDs can be provided by a variety of junction types, including tunnel junctions, micro-bridges, and proximity weak links \cite{likharev1979}. 
In the latter, a short, non-intrinsically-superconducting element, such as a normal metallic wire allows for superconducting correlations to propagate between both superconducting reservoirs. 
Due to the wire's usually low normal state resistance, following the resistively and capacitively shunted Josephson junction model, the quantum phase dynamics in such superconductor-normal metal-superconductor (SNS) Josephson junctions are inherently overdamped \elke{\cite{tinkham1996dover}}. This ensures that upon a QPS, the quantum phase only evolves towards the nearest neighboring potential valley (\mbox{Fig.} \ref{fig:schematics}c). This releases an energy 
\begin{equation}
\Delta U=\frac{\Phi_0}{2\pi L}\int \Phi_x \,d\varphi,
\label{eq:DeltaU} 
\end{equation}
which depends on the magnitude of the quantum phase slip. In the large $\beta$ limit, the phase jumps by about $2\pi$, and $\Delta U\approx I_c\Phi_0$. 

Our device consists of a SQUID with a single Josephson junction, provided by an SNS weak link of length $520$ nm (\mbox{Fig.} \ref{fig:static}a). Here, the superconducting circuit parts are made of aluminum, while the metal N is made of copper. 
While the loop is grounded, the center of N is further contacted by another superconducting finger through a tunnel junction (\mbox{Fig.} \ref{fig:static}a), with normal state resistance $R_T\approx7$ k$\Omega$.
This SQUID variant was named SQUIPT \cite{giazotto2010}, where PT stands for proximity transistor.
Our SQUIPT was designed to be in the hysteretic regime. For this, we took into account the geometric and kinetic inductance contributions, leading here to $L\approx630\pm50$ pH. The SNS junction critical current $I_c$ cannot be determined independently in this device, but is expected from similar SNS devices in a current-biased geometry to be of a few $\mu$A \cite{courtois2008,angers2008,dutta2020}. From this, values of the screening parameter $\beta\sim 10$ can be anticipated, in good agreement with experiments, as discussed below.

Applying a dc voltage $V_b$ to the tunnel junction, we can  perform tunnel spectroscopy by measuring its differential conductance $G(V_b)$. Here, $G$ is not read out, as usual, by a low-frequency transport measurement, but by a radio-frequency (RF) technique, using a superconducting $LC$ resonator with resonant frequency 575 MHz \cite{schmidt2003,gasparinetti2015,karimi2018}. 
By embedding the tunnel junction in parallel to the resonator, and for a fixed incident RF power $P_{\rm in}$, changes in $G$ can be read out by their effect on the transmitted power at resonance $P_{\rm out}$, which we record after cryogenic amplification (details in \mbox{Supp. Info.} file).
This has the paramount advantage of allowing for extremely rapid measurements, limited by the resonator bandwidth, of about 10 MHz here. 

Figure \ref{fig:static}b shows measurements of $P_{\rm out}(V_b)$ at two temperatures.
Several characteristic spectroscopic features stand out, in particular (i) a spectroscopic gap of total width 480 $\mu$V, (ii) subgap resonances near $\pm 190\,\mu$V visible only at 400 mK, and (iii) three low-energy resonances at 0 and $\pm15\,\mu$V, respectively.
Keeping in mind that the tunnel junction connects an intrinsic superconductor with gap $\Delta$ and a proximized metal with a (smaller) induced gap $E_g$, we can evaluate the total spectroscopic gap as $2(\Delta+E_g)/e$. 
At intermediate temperatures, thermally activated conductance resonances occur at $\pm|\Delta-E_g|/e$, which we identify as feature (ii).
From the two above relations we find a gap $\Delta=210\pm5\,\mu $eV in the superconducting probe electrode at 400 mK (225 $\mu$eV at 50 mK), a typical value in nanostructured aluminum. We further extract $E_g= 29\,\mu$eV, in good agreement with theoretical estimates (details in the \mbox{Supp. Info.} file).
Eventually, the conductance resonance at $V_b=0$ (feature (iii)) is a signature of the Josephson coupling across the tunnel junction, which was purposely designed  to have an intermediate transparency \cite{karimi2018}. The satellite peaks at $\pm 15\,\mu$V are probably the signature of inelastic Cooper pair tunneling \cite{karimi2018} and are not of central relevance to this work.

As visible in \mbox{Fig.} \ref{fig:static}b,c and already discussed in detail in \cite{karimi2018,karimi2020pra, karimi2020natcomm}, the zero-bias conductance of the tunnel junction, and therefore $P_{\rm out}(V_b=0)$, is a sensitive probe of the electron temperature $T$ in N. 
Accordingly, we set $V_b=0$ in the remainder of this work and use $P_{\rm out}$ for both static and dynamic electron thermometry, after initial calibration under equilibrium conditions (\mbox{Fig.} \ref{fig:static}c).
However, and in contrast to previous work \cite{karimi2018,karimi2020pra, karimi2020natcomm} in which N was not subject to a phase drop, in the present device the tunnel junction conductance and thus $P_{\rm out}$ are clearly also a function of the phase $\varphi$ across the SNS junction. 
This is seen in \mbox{Fig.} \ref{fig:static}d, where sweeping the applied flux translates into a phase variation via \mbox{Eq.} (\ref{eq:phi-flux}). As $\varphi \to \pi/2$, the decrease of $E_g$ \cite{zhou1998,lesueur2008} entails a rapidly shrinking Josephson energy $E_J$ of the tunnel junction, and thus a decrease of $G(0)$, that is, an increase of $P_{\rm out}$. 
When the switching point in the $\Phi_x(\varphi)$ relation is reached, $\varphi$ suddenly relaxes to a smaller value (modulo $2\pi$), restoring $E_J$ and thus leading to an abrupt drop in $P_{\rm out}$. 
As expected, the same pattern is repeated with period $\Phi_0$ in the applied flux and mirror symmetric under inversion of the sweep direction \cite{ligato2021}. 
At higher temperatures, $\beta\propto I_c$ decreases and the modulation amplitude of $P_{\rm out}(\Phi_x)$ shrinks, while the $\Phi_0$-periodicity of the signal is preserved.

\begin{figure*}
    \centering
    %\vspace{-2cm}
    \includegraphics[width=0.95\textwidth]{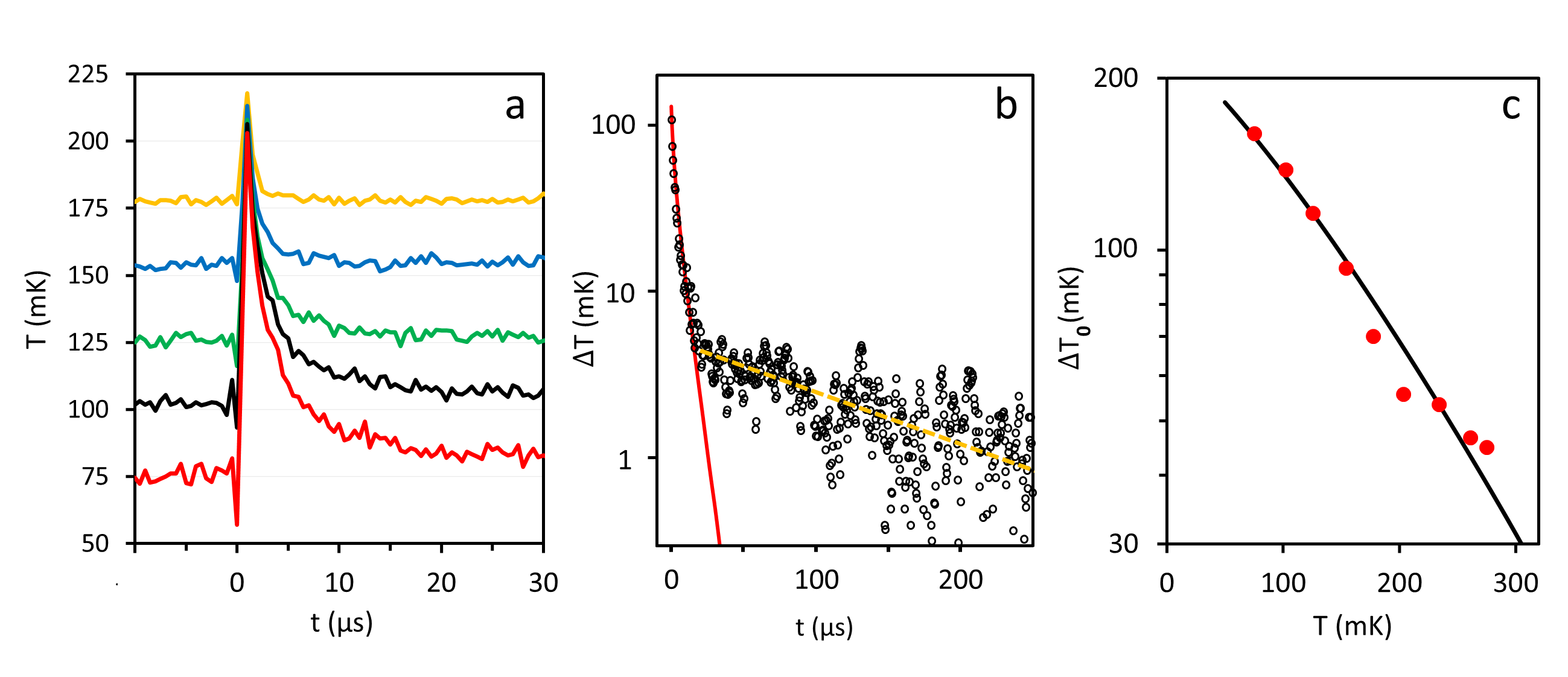}
    \vspace{-0.7cm}
    \caption{{Heat relaxation dynamics after a QPS.} (a) Time-resolved electron temperature in the absorber, at different starting temperatures set by the cryostat bath, following a $70$-ns flux pulse at $t=0$. The data sampling rate is 2 MHz, the data shown are the result of averaging over $10^5$ pulses, at a repetition rate of 2 kHz. (b) Return to equilibrium $\Delta T (t)$ at 100 mK, following a flux pulse. Same data as the black curve in (a), but in a semi-log-scale representation and over a wider time window. The red line is a calculation based on the model discussed in the text. The dashed line is an exponential fit (with time constant $\tau \approx 140\,\mu$s) to the long-term relaxation, evidencing the presence of a second slowly relaxing bath. (c) Magnitude of the initial temperature rise $\Delta T_0$ at $t=0$, determined by the fit as shown in (b) (bullets). The solid line is a calculation (see text).}
    \label{fig:relaxation}
\end{figure*}

For a quantitative understanding of the RF-SQUIPT, we use the quasi-classical Usadel equations  \cite{LarkinOvchinnikov1968,Usadel1970,zhou1998,Belzig1999}, with a single consistent set of microscopic parameters, described in detail in the \mbox{Supp. Info.} file. The density of states in N is known to display a minigap, which depends approximately on $\varphi$ like $E_g(\varphi)= E_g(0)\, |\cos(\varphi/2)|$ \cite{zhou1998}. The tunnel junction connecting the condensate in N to the superconducting probe electrode has a Josephson energy $E_J(\varphi,T)$ and thus a zero-bias conductance $G(\varphi,T)$, which can be drawn back analytically to $E_g(\varphi)$ and the critical current $I_c$ of the SNS junction.
In combination with \mbox{Eq.} (\ref{eq:phi-flux}) and the relation between $G$ and $P_{\rm out}$, the calculation provides an accurate description of the applied-flux dependence of the RF signal (\mbox{Fig.} \ref{fig:static}d), the only adjustable parameter being the magnitude of $\beta$. 

The temperature dependence of $\beta$ extracted from the data in Fig. \ref{fig:static}d is plotted in \mbox{Fig.} \ref{fig:static}e, following the trend expected for $I_c(T)$ in an SNS junction. The solid line shows the calculation from the same model as above, yielding a $5.9\,\mu$A zero-temperature critical current in the SNS junction. The parameters entering the calculation, and in particular the minigap, are determined independently using the tunnel spectra (\mbox{Fig.} \ref{fig:static}b).
We attribute the slight underestimation of $\beta$ by theory above 300 mK to the temperature dependence of the kinetic inductance, which was not accounted for. 

The data discussed so far, and summarized in \mbox{Fig.} \ref{fig:static}, provide a consistent physical understanding of the RF-SQUIPT under time-averaged, and thus isothermal conditions. We now move to the time-resolved response $P_{\rm out}(t)$, which displays the calorimetric signature of the heat deposited by individual QPS events. 
In addition to the static flux bias, we apply a time-dependent (square-wave or pulsed) flux, by passing a current $I_{\rm RFL}(t)$ through the superconducting rapid flux line (RFL), visible on the left side of the SQUIPT in \mbox{Fig.} \ref{fig:static}a. The instantaneous flux bias is $\Phi_x(t)=M\,I_{\rm RFL}(t)$, with a mutual inductance $M=12.1$ pH further discussed in the \mbox{Supp. Info.} file. In order to increase the signal-to-noise ratio, we average the resonator response over a large number of periodically generated identical $70$ ns flux pulses.

As long as the amplitude of the flux pulses does not exceed the threshold leading to a QPS, $P_{\rm out}$ follows changes in $\Phi_x(t)$ instantaneously (not shown, see \mbox{Supp. Info.} file for details). However, as soon as the threshold to instability of the flux state is overcome, $P_{\rm out}(t)$ displays in addition a slower relaxation to its novel equilibrium, which is indicative of thermalization. As evidenced from \mbox{Fig.} \ref{fig:relaxation}a, the relaxation dynamics after a flux pulse are strongly temperature dependent, as expected for instance from a dominantly electron-phonon (e-ph) coupling-driven thermalization after an initial heating event \cite{gasparinetti2015}. Above 300 mK, the relaxation times become too short ($<1\, \mu$s) to be measured.

The thermal dynamics can be described by a heat balance equation, 
basing on the standard assumption that variations of the absorber's internal energy $U$ are evacuated to a heat bath via electron-phonon coupling. In metallic nanostructures at low temperatures, this power is usually written ${\dot Q}_{e-ph}=\Sigma {\mathcal V} (T^5-T_0^5)$, where  $\Sigma=2\times10^{9}$ W\,K$^{-5}$\,m$^{-3}$ is the electron-phonon coupling constant in copper \cite{giazotto2006opportunities,viisanen2018}, and ${\mathcal V}\approx 8\times 10^{-21}$ m$^{-3}$ the geometrically estimated absorber volume. Given the rather small $E_g$ in the SNS junction, our calculations indicate that the proximity effect should only lead to negligibly small departures from the normal-state electron-phonon coupling and heat capacity $C$, at the experimental temperatures. 
Therefore we write $U= \gamma T^2/2$, using a reported $\gamma=71$ J\,m$^{-3}$\,K$^{-2}$ in nanostructured Cu \cite{viisanen2018}. 
%(instead of a tabulated $98$ J\,m$^{-3}$\,K$^{-2}$  in the bulk). 
As seen in \mbox{Fig.} \ref{fig:relaxation}a, the temperature increase after the initial heat pulse can be large compared to the starting temperature. The dynamical heat balance differential equation thus cannot be linearized in $\Delta T=T-T_b$, and must be solved numerically. The result at $T_b=100$ mK is shown in \mbox{Fig.} \ref{fig:relaxation}b (red line), where the only free parameter is the initial temperature increase $\Delta T_0$. The calculation follows closely the data during the first period of the thermal relaxation.
Interestingly, the initial fast decay is rapidly taken over by a much slower process, which was already reported for nanoscale Cu absorbers \cite{viisanen2018,wang2019}. This might be due to another heat reservoir, for instance in surface states of the Cu absorber.

The initial temperature increase $\Delta T_0$ after a flux pulse is plotted as a function of the starting temperature in \mbox{Fig.} \ref{fig:relaxation}c. Naturally, the temperature rise is highest when starting from the lowest base temperature, because at higher temperature both $C$ increases and $\Delta U \sim I_c\Phi_0$ decreases. For a quantitative modeling, one must bear in mind that a flux pulse induces necessarily two QPSs: one while ramping up the flux bias, and the second when returning to the initial state. 
We account for the fact that the return QPS occurs at a higher electronic temperature and thus produces less dissipation.
%If $\beta$ was to go below 1 after the first QPS, the return to the initial state could even be non-dissipative, which is however not the case in this device. 
The sudden temperature rise after the first QPS could also lead the sinusoidal potential energy landscape modulation (see \mbox{Fig.} \ref{fig:schematics}c) to collapse to the point that the reached valley also becomes unstable and the phase eventually jumps by $4\pi$, or more. This was actually observed in another device, which we describe in the \mbox{Supp. Info.} file, but does not apply in the sample described here. Finally, the quasi-classical Usadel formalism with the same set of microscopic parameters as previously, in combination with \mbox{Eqs.} (\ref{eq:phi-flux},\ref{eq:DeltaU}) and the known values of $\gamma$ and ${\mathcal V}$, describes very accurately (red line in \mbox{Fig.} \ref{fig:relaxation}c) the initial temperature rise $\Delta T_0$.

Our findings of microsecond-scale thermal relaxation following a QPS tracked in real time highlight both the paramount effect dissipation can produce in quantum circuits, and the potential of large-bandwidth electron thermometry for quantum thermodynamics in nanoelectronic circuits. 
The detection of a {\it quantum of dissipation} in a Josephson junction opens promising perspectives for future experiments. These would aim, for instance, at detecting the minute dissipation arising from a projective qubit measurement with emission of a single microwave photon \cite{pekola2022}, or the detection of the elusive axion with even smaller energy \cite{braine2020}. On the other hand, the proper measurement and control of dissipation in QPS-based circuits will allow for overcoming self-heating-limited device performances.

We acknowledge help from A. Th\'ery and T. Crozes. The samples were fabricated at the Nanofab platform at Institut N\'eel. This work received support from the European Union under the Marie Sklodowska-Curie Grant Agreement \mbox{No.} 766025 (QuESTech), from the Agence Nationale de la Recherche under the program “Investissements d’avenir” (ANR-15-IDEX-02), from the Laboratoire d'excellence LANEF (ANR-10-LABX-51-01), and from Deutsche Forschungsgemeinschaft (DFG; German Research Foundation) via SFB 1432 (Project \mbox{No.}
425217212).

%{\bf Author contributions} EG performed the experiments, with the help of PR. DM, EG and PR made the samples. DN and WB did the theoretical modelling. CBW conducted the research and wrote the manuscript. All authors contributed to the discussions and interpretation of the data.

%\tableofcontent
%\begin{thebibliography}{4}
%\section{References}
%\nocite{*}
\bibliography{mybib}% Produces the bibliography via BibTeX.

%merlin.mbs apsrev4-1.bst 2010-07-25 4.21a (PWD, AO, DPC) hacked
%Control: key (0)
%Control: author (0) dotless jnrlst
%Control: editor formatted (1) identically to author
%Control: production of article title (0) allowed
%Control: page (1) range
%Control: year (0) verbatim
%Control: production of eprint (0) enabled
\begin{thebibliography}{37}%
\makeatletter
\providecommand \@ifxundefined [1]{%
 \@ifx{#1\undefined}
}%
\providecommand \@ifnum [1]{%
 \ifnum #1\expandafter \@firstoftwo
 \else \expandafter \@secondoftwo
 \fi
}%
\providecommand \@ifx [1]{%
 \ifx #1\expandafter \@firstoftwo
 \else \expandafter \@secondoftwo
 \fi
}%
\providecommand \natexlab [1]{#1}%
\providecommand \enquote  [1]{``#1''}%
\providecommand \bibnamefont  [1]{#1}%
\providecommand \bibfnamefont [1]{#1}%
\providecommand \citenamefont [1]{#1}%
\providecommand \href@noop [0]{\@secondoftwo}%
\providecommand \href [0]{\begingroup \@sanitize@url \@href}%
\providecommand \@href[1]{\@@startlink{#1}\@@href}%
\providecommand \@@href[1]{\endgroup#1\@@endlink}%
\providecommand \@sanitize@url [0]{\catcode `\\12\catcode `\$12\catcode
  `\&12\catcode `\#12\catcode `\^12\catcode `\_12\catcode `\%12\relax}%
\providecommand \@@startlink[1]{}%
\providecommand \@@endlink[0]{}%
\providecommand \url  [0]{\begingroup\@sanitize@url \@url }%
\providecommand \@url [1]{\endgroup\@href {#1}{\urlprefix }}%
\providecommand \urlprefix  [0]{URL }%
\providecommand \Eprint [0]{\href }%
\providecommand \doibase [0]{http://dx.doi.org/}%
\providecommand \selectlanguage [0]{\@gobble}%
\providecommand \bibinfo  [0]{\@secondoftwo}%
\providecommand \bibfield  [0]{\@secondoftwo}%
\providecommand \translation [1]{[#1]}%
\providecommand \BibitemOpen [0]{}%
\providecommand \bibitemStop [0]{}%
\providecommand \bibitemNoStop [0]{.\EOS\space}%
\providecommand \EOS [0]{\spacefactor3000\relax}%
\providecommand \BibitemShut  [1]{\csname bibitem#1\endcsname}%
\let\auto@bib@innerbib\@empty
%</preamble>
\bibitem [{\citenamefont {Mooij}\ and\ \citenamefont
  {Harmans}(2005)}]{mooij2005}%
  \BibitemOpen
  \bibfield  {author} {\bibinfo {author} {\bibfnamefont {J.~E.}\ \bibnamefont
  {Mooij}}\ and\ \bibinfo {author} {\bibfnamefont {C.~J. P.~M.}\ \bibnamefont
  {Harmans}},\ }\bibfield  {title} {\enquote {\bibinfo {title} {Phase-slip flux
  qubits},}\ }\href@noop {} {\bibfield  {journal} {\bibinfo  {journal} {New J.
  Phys.}\ }\textbf {\bibinfo {volume} {7}},\ \bibinfo {pages} {219} (\bibinfo
  {year} {2005})}\BibitemShut {NoStop}%
\bibitem [{\citenamefont {Mooij}\ and\ \citenamefont
  {Nazarov}(2006)}]{mooij2006}%
  \BibitemOpen
  \bibfield  {author} {\bibinfo {author} {\bibfnamefont {J.~E.}\ \bibnamefont
  {Mooij}}\ and\ \bibinfo {author} {\bibfnamefont {Yu.~V.}\ \bibnamefont
  {Nazarov}},\ }\bibfield  {title} {\enquote {\bibinfo {title} {Superconducting
  nanowires as quantum phase-slip junctions},}\ }\href@noop {} {\bibfield
  {journal} {\bibinfo  {journal} {Nat. Phys.}\ }\textbf {\bibinfo {volume}
  {2}},\ \bibinfo {pages} {169--172} (\bibinfo {year} {2006})}\BibitemShut
  {NoStop}%
\bibitem [{\citenamefont {Liebermann}\ and\ \citenamefont
  {Wilhelm}(2016)}]{liebermann2016}%
  \BibitemOpen
  \bibfield  {author} {\bibinfo {author} {\bibfnamefont {P.~J.}\ \bibnamefont
  {Liebermann}}\ and\ \bibinfo {author} {\bibfnamefont {F.~K.}\ \bibnamefont
  {Wilhelm}},\ }\bibfield  {title} {\enquote {\bibinfo {title} {Optimal qubit
  control using single-flux quantum pulses},}\ }\href@noop {} {\bibfield
  {journal} {\bibinfo  {journal} {Phys. Rev. Appl.}\ }\textbf {\bibinfo
  {volume} {6}},\ \bibinfo {pages} {024022} (\bibinfo {year}
  {2016})}\BibitemShut {NoStop}%
\bibitem [{\citenamefont {Leonard}\ \emph {et~al.}(2019)\citenamefont
  {Leonard}, \citenamefont {Beck}, \citenamefont {Nelson}, \citenamefont
  {Christensen}, \citenamefont {Thorbeck}, \citenamefont {Howington},
  \citenamefont {Opremcak}, \citenamefont {Pechenezhskiy}, \citenamefont
  {Dodge}, \citenamefont {Dupuis}, \citenamefont {Hutchings}, \citenamefont
  {Ku}, \citenamefont {Schlenker}, \citenamefont {Suttle}, \citenamefont
  {Wilen}, \citenamefont {Zhu}, \citenamefont {Vavilov}, \citenamefont
  {Plourde},\ and\ \citenamefont {McDermott}}]{leonard_digital_2019}%
  \BibitemOpen
  \bibfield  {author} {\bibinfo {author} {\bibfnamefont {E.}~\bibnamefont
  {Leonard}}, \bibinfo {author} {\bibfnamefont {M.~A.}\ \bibnamefont {Beck}},
  \bibinfo {author} {\bibfnamefont {J.}~\bibnamefont {Nelson}}, \bibinfo
  {author} {\bibfnamefont {B.~G.}\ \bibnamefont {Christensen}}, \bibinfo
  {author} {\bibfnamefont {T.}~\bibnamefont {Thorbeck}}, \bibinfo {author}
  {\bibfnamefont {C.}~\bibnamefont {Howington}}, \bibinfo {author}
  {\bibfnamefont {A.}~\bibnamefont {Opremcak}}, \bibinfo {author}
  {\bibfnamefont {I.~V.}\ \bibnamefont {Pechenezhskiy}}, \bibinfo {author}
  {\bibfnamefont {K.}~\bibnamefont {Dodge}}, \bibinfo {author} {\bibfnamefont
  {N.~P.}\ \bibnamefont {Dupuis}}, \bibinfo {author} {\bibfnamefont {M.~D.}\
  \bibnamefont {Hutchings}}, \bibinfo {author} {\bibfnamefont {J.}~\bibnamefont
  {Ku}}, \bibinfo {author} {\bibfnamefont {F.}~\bibnamefont {Schlenker}},
  \bibinfo {author} {\bibfnamefont {J.}~\bibnamefont {Suttle}}, \bibinfo
  {author} {\bibfnamefont {C.}~\bibnamefont {Wilen}}, \bibinfo {author}
  {\bibfnamefont {S.}~\bibnamefont {Zhu}}, \bibinfo {author} {\bibfnamefont
  {M.~G.}\ \bibnamefont {Vavilov}}, \bibinfo {author} {\bibfnamefont
  {B.~L.~T.}\ \bibnamefont {Plourde}}, \ and\ \bibinfo {author} {\bibfnamefont
  {R.}~\bibnamefont {McDermott}},\ }\bibfield  {title} {\enquote {\bibinfo
  {title} {Digital {Coherent} {Control} of a {Superconducting} {Qubit}},}\
  }\href@noop {} {\bibfield  {journal} {\bibinfo  {journal} {Phys. Rev. Appl.}\
  }\textbf {\bibinfo {volume} {11}},\ \bibinfo {pages} {014009} (\bibinfo
  {year} {2019})}\BibitemShut {NoStop}%
\bibitem [{\citenamefont {Howington}\ \emph {et~al.}(2019)\citenamefont
  {Howington}, \citenamefont {Opremcak}, \citenamefont {McDermott},
  \citenamefont {Kirichenko}, \citenamefont {Mukhanov},\ and\ \citenamefont
  {Plourde}}]{howington_interfacing_2019}%
  \BibitemOpen
  \bibfield  {author} {\bibinfo {author} {\bibfnamefont {C.}~\bibnamefont
  {Howington}}, \bibinfo {author} {\bibfnamefont {A.}~\bibnamefont {Opremcak}},
  \bibinfo {author} {\bibfnamefont {R.}~\bibnamefont {McDermott}}, \bibinfo
  {author} {\bibfnamefont {A.}~\bibnamefont {Kirichenko}}, \bibinfo {author}
  {\bibfnamefont {O.~A.}\ \bibnamefont {Mukhanov}}, \ and\ \bibinfo {author}
  {\bibfnamefont {B.~L.~T.}\ \bibnamefont {Plourde}},\ }\bibfield  {title}
  {\enquote {\bibinfo {title} {Interfacing {Superconducting} {Qubits} {With}
  {Cryogenic} {Logic}: {Readout}},}\ }\href@noop {} {\bibfield  {journal}
  {\bibinfo  {journal} {IEEE Transactions on Applied Superconductivity}\
  }\textbf {\bibinfo {volume} {29}},\ \bibinfo {pages} {1--5} (\bibinfo {year}
  {2019})}\BibitemShut {NoStop}%
\bibitem [{\citenamefont {McDermott}\ \emph {et~al.}(2018)\citenamefont
  {McDermott}, \citenamefont {Vavilov}, \citenamefont {Plourde}, \citenamefont
  {Wilhelm}, \citenamefont {Liebermann}, \citenamefont {Mukhanov},\ and\
  \citenamefont {Ohki}}]{mcdermott_quantumclassical_2018}%
  \BibitemOpen
  \bibfield  {author} {\bibinfo {author} {\bibfnamefont {R.}~\bibnamefont
  {McDermott}}, \bibinfo {author} {\bibfnamefont {M.~G.}\ \bibnamefont
  {Vavilov}}, \bibinfo {author} {\bibfnamefont {B.~L.~T.}\ \bibnamefont
  {Plourde}}, \bibinfo {author} {\bibfnamefont {F.~K.}\ \bibnamefont
  {Wilhelm}}, \bibinfo {author} {\bibfnamefont {P.~J.}\ \bibnamefont
  {Liebermann}}, \bibinfo {author} {\bibfnamefont {O.~A.}\ \bibnamefont
  {Mukhanov}}, \ and\ \bibinfo {author} {\bibfnamefont {T.~A.}\ \bibnamefont
  {Ohki}},\ }\bibfield  {title} {\enquote {\bibinfo {title}
  {Quantum–classical interface based on single flux quantum digital logic},}\
  }\href@noop {} {\bibfield  {journal} {\bibinfo  {journal} {Quantum Science
  and Technology}\ }\textbf {\bibinfo {volume} {3}},\ \bibinfo {pages} {024004}
  (\bibinfo {year} {2018})}\BibitemShut {NoStop}%
\bibitem [{\citenamefont {Chiorescu}\ \emph {et~al.}(2003)\citenamefont
  {Chiorescu}, \citenamefont {Nakamura}, \citenamefont {Harmans},\ and\
  \citenamefont {Mooij}}]{chiorescu2003}%
  \BibitemOpen
  \bibfield  {author} {\bibinfo {author} {\bibfnamefont {I.}~\bibnamefont
  {Chiorescu}}, \bibinfo {author} {\bibfnamefont {Y.}~\bibnamefont {Nakamura}},
  \bibinfo {author} {\bibfnamefont {C.~J.~P.}\ \bibnamefont {Harmans}}, \ and\
  \bibinfo {author} {\bibfnamefont {J.~E.}\ \bibnamefont {Mooij}},\ }\bibfield
  {title} {\enquote {\bibinfo {title} {Coherent quantum dynamics of a
  superconducting flux qubit},}\ }\href@noop {} {\bibfield  {journal} {\bibinfo
   {journal} {Science}\ }\textbf {\bibinfo {volume} {299}},\ \bibinfo {pages}
  {1869--1871} (\bibinfo {year} {2003})}\BibitemShut {NoStop}%
\bibitem [{\citenamefont {Yan}\ \emph {et~al.}(2016)\citenamefont {Yan},
  \citenamefont {Gustavsson}, \citenamefont {Kamal}, \citenamefont {Birenbaum},
  \citenamefont {Sears}, \citenamefont {Hover}, \citenamefont {Gudmundsen},
  \citenamefont {Rosenberg}, \citenamefont {Samach}, \citenamefont {Weber}
  \emph {et~al.}}]{yan2016}%
  \BibitemOpen
  \bibfield  {author} {\bibinfo {author} {\bibfnamefont {F.}~\bibnamefont
  {Yan}}, \bibinfo {author} {\bibfnamefont {S.}~\bibnamefont {Gustavsson}},
  \bibinfo {author} {\bibfnamefont {A.}~\bibnamefont {Kamal}}, \bibinfo
  {author} {\bibfnamefont {J.}~\bibnamefont {Birenbaum}}, \bibinfo {author}
  {\bibfnamefont {A.~P.}\ \bibnamefont {Sears}}, \bibinfo {author}
  {\bibfnamefont {D.}~\bibnamefont {Hover}}, \bibinfo {author} {\bibfnamefont
  {T.~J.}\ \bibnamefont {Gudmundsen}}, \bibinfo {author} {\bibfnamefont
  {D.}~\bibnamefont {Rosenberg}}, \bibinfo {author} {\bibfnamefont
  {G.}~\bibnamefont {Samach}}, \bibinfo {author} {\bibfnamefont
  {S.}~\bibnamefont {Weber}},  \emph {et~al.},\ }\bibfield  {title} {\enquote
  {\bibinfo {title} {The flux qubit revisited to enhance coherence and
  reproducibility},}\ }\href@noop {} {\bibfield  {journal} {\bibinfo  {journal}
  {Nat. Commun.}\ }\textbf {\bibinfo {volume} {7}},\ \bibinfo {pages} {1--9}
  (\bibinfo {year} {2016})}\BibitemShut {NoStop}%
\bibitem [{\citenamefont {Manucharyan}\ \emph {et~al.}(2009)\citenamefont
  {Manucharyan}, \citenamefont {Koch}, \citenamefont {Glazman},\ and\
  \citenamefont {Devoret}}]{Manucharyan2009}%
  \BibitemOpen
  \bibfield  {author} {\bibinfo {author} {\bibfnamefont {V.~E.}\ \bibnamefont
  {Manucharyan}}, \bibinfo {author} {\bibfnamefont {J.}~\bibnamefont {Koch}},
  \bibinfo {author} {\bibfnamefont {L.~I.}\ \bibnamefont {Glazman}}, \ and\
  \bibinfo {author} {\bibfnamefont {M.~H.}\ \bibnamefont {Devoret}},\
  }\bibfield  {title} {\enquote {\bibinfo {title} {Fluxonium: Single
  {C}ooper-pair circuit free of charge offsets},}\ }\href@noop {} {\bibfield
  {journal} {\bibinfo  {journal} {Science}\ }\textbf {\bibinfo {volume}
  {326}},\ \bibinfo {pages} {113} (\bibinfo {year} {2009})}\BibitemShut
  {NoStop}%
\bibitem [{\citenamefont {Nguyen}\ \emph {et~al.}(2019)\citenamefont {Nguyen},
  \citenamefont {Lin}, \citenamefont {Somoroff}, \citenamefont {Mencia},
  \citenamefont {Grabon},\ and\ \citenamefont {Manucharyan}}]{nguyen2019}%
  \BibitemOpen
  \bibfield  {author} {\bibinfo {author} {\bibfnamefont {L.~B.}\ \bibnamefont
  {Nguyen}}, \bibinfo {author} {\bibfnamefont {Y.-H.}\ \bibnamefont {Lin}},
  \bibinfo {author} {\bibfnamefont {A.}~\bibnamefont {Somoroff}}, \bibinfo
  {author} {\bibfnamefont {R.}~\bibnamefont {Mencia}}, \bibinfo {author}
  {\bibfnamefont {N.}~\bibnamefont {Grabon}}, \ and\ \bibinfo {author}
  {\bibfnamefont {V.~E}\ \bibnamefont {Manucharyan}},\ }\bibfield  {title}
  {\enquote {\bibinfo {title} {High-coherence fluxonium qubit},}\ }\href@noop
  {} {\bibfield  {journal} {\bibinfo  {journal} {Phys. Rev. X}\ }\textbf
  {\bibinfo {volume} {9}},\ \bibinfo {pages} {041041} (\bibinfo {year}
  {2019})}\BibitemShut {NoStop}%
\bibitem [{\citenamefont {Astafiev}\ \emph {et~al.}(2012)\citenamefont
  {Astafiev}, \citenamefont {Ioffe}, \citenamefont {Kafanov}, \citenamefont
  {Pashkin}, \citenamefont {Arutyunov}, \citenamefont {Shahar}, \citenamefont
  {Cohen},\ and\ \citenamefont {Tsai}}]{astafiev2012}%
  \BibitemOpen
  \bibfield  {author} {\bibinfo {author} {\bibfnamefont {O.~V.}\ \bibnamefont
  {Astafiev}}, \bibinfo {author} {\bibfnamefont {L.~B.}\ \bibnamefont {Ioffe}},
  \bibinfo {author} {\bibfnamefont {S.}~\bibnamefont {Kafanov}}, \bibinfo
  {author} {\bibfnamefont {Y.~A.}\ \bibnamefont {Pashkin}}, \bibinfo {author}
  {\bibfnamefont {K.~Y.}\ \bibnamefont {Arutyunov}}, \bibinfo {author}
  {\bibfnamefont {D.}~\bibnamefont {Shahar}}, \bibinfo {author} {\bibfnamefont
  {O.}~\bibnamefont {Cohen}}, \ and\ \bibinfo {author} {\bibfnamefont {J.~S.}\
  \bibnamefont {Tsai}},\ }\bibfield  {title} {\enquote {\bibinfo {title}
  {Coherent quantum phase slip},}\ }\href@noop {} {\bibfield  {journal}
  {\bibinfo  {journal} {Nature}\ }\textbf {\bibinfo {volume} {484}},\ \bibinfo
  {pages} {355--358} (\bibinfo {year} {2012})}\BibitemShut {NoStop}%
\bibitem [{\citenamefont {Golubev}\ and\ \citenamefont
  {Zaikin}(2001)}]{golubev2001}%
  \BibitemOpen
  \bibfield  {author} {\bibinfo {author} {\bibfnamefont {D.~S.}\ \bibnamefont
  {Golubev}}\ and\ \bibinfo {author} {\bibfnamefont {A.~D.}\ \bibnamefont
  {Zaikin}},\ }\bibfield  {title} {\enquote {\bibinfo {title} {Quantum
  tunneling of the order parameter in superconducting nanowires},}\ }\href@noop
  {} {\bibfield  {journal} {\bibinfo  {journal} {Phys. Rev. B}\ }\textbf
  {\bibinfo {volume} {64}},\ \bibinfo {pages} {014504} (\bibinfo {year}
  {2001})}\BibitemShut {NoStop}%
\bibitem [{\citenamefont {Sahu}\ \emph {et~al.}(2009)\citenamefont {Sahu},
  \citenamefont {Bae}, \citenamefont {Rogachev}, \citenamefont {Pekker},
  \citenamefont {Wei}, \citenamefont {Shah}, \citenamefont {Goldbart},\ and\
  \citenamefont {Bezryadin}}]{sahu2009}%
  \BibitemOpen
  \bibfield  {author} {\bibinfo {author} {\bibfnamefont {M.}~\bibnamefont
  {Sahu}}, \bibinfo {author} {\bibfnamefont {M.}~\bibnamefont {Bae}}, \bibinfo
  {author} {\bibfnamefont {A.}~\bibnamefont {Rogachev}}, \bibinfo {author}
  {\bibfnamefont {D.}~\bibnamefont {Pekker}}, \bibinfo {author} {\bibfnamefont
  {T.-C.}\ \bibnamefont {Wei}}, \bibinfo {author} {\bibfnamefont
  {N.}~\bibnamefont {Shah}}, \bibinfo {author} {\bibfnamefont {P.~M.}\
  \bibnamefont {Goldbart}}, \ and\ \bibinfo {author} {\bibfnamefont
  {A.}~\bibnamefont {Bezryadin}},\ }\bibfield  {title} {\enquote {\bibinfo
  {title} {Individual topological tunnelling events of a quantum field probed
  through their macroscopic consequences},}\ }\href@noop {} {\bibfield
  {journal} {\bibinfo  {journal} {Nat. Phys.}\ }\textbf {\bibinfo {volume}
  {5}},\ \bibinfo {pages} {503--508} (\bibinfo {year} {2009})}\BibitemShut
  {NoStop}%
\bibitem [{\citenamefont {Tinkham}(1996)}]{tinkham1996dover}%
  \BibitemOpen
  \bibfield  {author} {\bibinfo {author} {\bibfnamefont {M.}~\bibnamefont
  {Tinkham}},\ }\href@noop {} {\emph {\bibinfo {title} {Introduction to
  Superconductivity}}}\ (\bibinfo  {publisher} {Dover, New York, 2nd edition},\
  \bibinfo {year} {1996})\BibitemShut {NoStop}%
\bibitem [{\citenamefont {van Der~Wal}\ \emph {et~al.}(2000)\citenamefont {van
  Der~Wal}, \citenamefont {Ter~Haar}, \citenamefont {Wilhelm}, \citenamefont
  {Schouten}, \citenamefont {Harmans}, \citenamefont {Orlando}, \citenamefont
  {Lloyd},\ and\ \citenamefont {Mooij}}]{vanderwal2000}%
  \BibitemOpen
  \bibfield  {author} {\bibinfo {author} {\bibfnamefont {C.~H.}\ \bibnamefont
  {van Der~Wal}}, \bibinfo {author} {\bibfnamefont {A.~C.~J.}\ \bibnamefont
  {Ter~Haar}}, \bibinfo {author} {\bibfnamefont {F.~K.}\ \bibnamefont
  {Wilhelm}}, \bibinfo {author} {\bibfnamefont {R.~N.}\ \bibnamefont
  {Schouten}}, \bibinfo {author} {\bibfnamefont {C.~J. P.~M.}\ \bibnamefont
  {Harmans}}, \bibinfo {author} {\bibfnamefont {T.~P.}\ \bibnamefont
  {Orlando}}, \bibinfo {author} {\bibfnamefont {S.}~\bibnamefont {Lloyd}}, \
  and\ \bibinfo {author} {\bibfnamefont {J.~E.}\ \bibnamefont {Mooij}},\
  }\bibfield  {title} {\enquote {\bibinfo {title} {Quantum superposition of
  macroscopic persistent-current states},}\ }\href@noop {} {\bibfield
  {journal} {\bibinfo  {journal} {Science}\ }\textbf {\bibinfo {volume}
  {290}},\ \bibinfo {pages} {773--777} (\bibinfo {year} {2000})}\BibitemShut
  {NoStop}%
\bibitem [{\citenamefont {Friedman}\ \emph {et~al.}(2000)\citenamefont
  {Friedman}, \citenamefont {Patel}, \citenamefont {Chen}, \citenamefont
  {Tolpygo},\ and\ \citenamefont {Lukens}}]{friedman2000}%
  \BibitemOpen
  \bibfield  {author} {\bibinfo {author} {\bibfnamefont {J.~R.}\ \bibnamefont
  {Friedman}}, \bibinfo {author} {\bibfnamefont {V.}~\bibnamefont {Patel}},
  \bibinfo {author} {\bibfnamefont {W.}~\bibnamefont {Chen}}, \bibinfo {author}
  {\bibfnamefont {S.~K.}\ \bibnamefont {Tolpygo}}, \ and\ \bibinfo {author}
  {\bibfnamefont {J.~E.}\ \bibnamefont {Lukens}},\ }\bibfield  {title}
  {\enquote {\bibinfo {title} {Quantum superposition of distinct macroscopic
  states},}\ }\href@noop {} {\bibfield  {journal} {\bibinfo  {journal}
  {Nature}\ }\textbf {\bibinfo {volume} {406}},\ \bibinfo {pages} {43--46}
  (\bibinfo {year} {2000})}\BibitemShut {NoStop}%
\bibitem [{\citenamefont {Likharev}(1979)}]{likharev1979}%
  \BibitemOpen
  \bibfield  {author} {\bibinfo {author} {\bibfnamefont {K.~K.}\ \bibnamefont
  {Likharev}},\ }\bibfield  {title} {\enquote {\bibinfo {title}
  {Superconducting weak links},}\ }\href@noop {} {\bibfield  {journal}
  {\bibinfo  {journal} {Rev. Mod. Phys.}\ }\textbf {\bibinfo {volume} {51}},\
  \bibinfo {pages} {101} (\bibinfo {year} {1979})}\BibitemShut {NoStop}%
\bibitem [{\citenamefont {Giazotto}\ \emph {et~al.}(2010)\citenamefont
  {Giazotto}, \citenamefont {Peltonen}, \citenamefont {Meschke},\ and\
  \citenamefont {Pekola}}]{giazotto2010}%
  \BibitemOpen
  \bibfield  {author} {\bibinfo {author} {\bibfnamefont {F.}~\bibnamefont
  {Giazotto}}, \bibinfo {author} {\bibfnamefont {J.~T.}\ \bibnamefont
  {Peltonen}}, \bibinfo {author} {\bibfnamefont {M.}~\bibnamefont {Meschke}}, \
  and\ \bibinfo {author} {\bibfnamefont {J.~P}\ \bibnamefont {Pekola}},\
  }\bibfield  {title} {\enquote {\bibinfo {title} {Superconducting quantum
  interference proximity transistor},}\ }\href@noop {} {\bibfield  {journal}
  {\bibinfo  {journal} {Nat. Phys.}\ }\textbf {\bibinfo {volume} {6}},\
  \bibinfo {pages} {254--259} (\bibinfo {year} {2010})}\BibitemShut {NoStop}%
\bibitem [{\citenamefont {Courtois}\ \emph {et~al.}(2008)\citenamefont
  {Courtois}, \citenamefont {Meschke}, \citenamefont {Peltonen},\ and\
  \citenamefont {Pekola}}]{courtois2008}%
  \BibitemOpen
  \bibfield  {author} {\bibinfo {author} {\bibfnamefont {H.}~\bibnamefont
  {Courtois}}, \bibinfo {author} {\bibfnamefont {M.}~\bibnamefont {Meschke}},
  \bibinfo {author} {\bibfnamefont {J.~T.}\ \bibnamefont {Peltonen}}, \ and\
  \bibinfo {author} {\bibfnamefont {J.~P}\ \bibnamefont {Pekola}},\ }\bibfield
  {title} {\enquote {\bibinfo {title} {Origin of hysteresis in a proximity
  {J}osephson junction},}\ }\href@noop {} {\bibfield  {journal} {\bibinfo
  {journal} {Phys. Rev. Lett.}\ }\textbf {\bibinfo {volume} {101}},\ \bibinfo
  {pages} {067002} (\bibinfo {year} {2008})}\BibitemShut {NoStop}%
\bibitem [{\citenamefont {Angers}\ \emph {et~al.}(2008)\citenamefont {Angers},
  \citenamefont {Chiodi}, \citenamefont {Montambaux}, \citenamefont {Ferrier},
  \citenamefont {Gu{\'e}ron}, \citenamefont {Bouchiat},\ and\ \citenamefont
  {Cuevas}}]{angers2008}%
  \BibitemOpen
  \bibfield  {author} {\bibinfo {author} {\bibfnamefont {L.}~\bibnamefont
  {Angers}}, \bibinfo {author} {\bibfnamefont {F.}~\bibnamefont {Chiodi}},
  \bibinfo {author} {\bibfnamefont {G.}~\bibnamefont {Montambaux}}, \bibinfo
  {author} {\bibfnamefont {M.}~\bibnamefont {Ferrier}}, \bibinfo {author}
  {\bibfnamefont {S.}~\bibnamefont {Gu{\'e}ron}}, \bibinfo {author}
  {\bibfnamefont {H.}~\bibnamefont {Bouchiat}}, \ and\ \bibinfo {author}
  {\bibfnamefont {J.~C.}\ \bibnamefont {Cuevas}},\ }\bibfield  {title}
  {\enquote {\bibinfo {title} {Proximity dc squids in the long-junction
  limit},}\ }\href@noop {} {\bibfield  {journal} {\bibinfo  {journal} {Phys.
  Rev. B}\ }\textbf {\bibinfo {volume} {77}},\ \bibinfo {pages} {165408}
  (\bibinfo {year} {2008})}\BibitemShut {NoStop}%
\bibitem [{\citenamefont {Dutta}\ \emph {et~al.}(2020)\citenamefont {Dutta},
  \citenamefont {Majidi}, \citenamefont {Talarico}, \citenamefont {Lo~Gullo},
  \citenamefont {Courtois},\ and\ \citenamefont {Winkelmann}}]{dutta2020}%
  \BibitemOpen
  \bibfield  {author} {\bibinfo {author} {\bibfnamefont {B.}~\bibnamefont
  {Dutta}}, \bibinfo {author} {\bibfnamefont {D.}~\bibnamefont {Majidi}},
  \bibinfo {author} {\bibfnamefont {N.~W.}\ \bibnamefont {Talarico}}, \bibinfo
  {author} {\bibfnamefont {N.}~\bibnamefont {Lo~Gullo}}, \bibinfo {author}
  {\bibfnamefont {H.}~\bibnamefont {Courtois}}, \ and\ \bibinfo {author}
  {\bibfnamefont {C.~B.}\ \bibnamefont {Winkelmann}},\ }\bibfield  {title}
  {\enquote {\bibinfo {title} {Single-quantum-dot heat valve},}\ }\href@noop {}
  {\bibfield  {journal} {\bibinfo  {journal} {Phys. Rev. Lett.}\ }\textbf
  {\bibinfo {volume} {125}},\ \bibinfo {pages} {237701} (\bibinfo {year}
  {2020})}\BibitemShut {NoStop}%
\bibitem [{\citenamefont {Schmidt}\ \emph {et~al.}(2003)\citenamefont
  {Schmidt}, \citenamefont {Yung},\ and\ \citenamefont
  {Cleland}}]{schmidt2003}%
  \BibitemOpen
  \bibfield  {author} {\bibinfo {author} {\bibfnamefont {D.~R.}\ \bibnamefont
  {Schmidt}}, \bibinfo {author} {\bibfnamefont {C.~S.}\ \bibnamefont {Yung}}, \
  and\ \bibinfo {author} {\bibfnamefont {A.~N.}\ \bibnamefont {Cleland}},\
  }\bibfield  {title} {\enquote {\bibinfo {title} {Nanoscale radio-frequency
  thermometry},}\ }\href@noop {} {\bibfield  {journal} {\bibinfo  {journal}
  {Appl. Phys. Lett.}\ }\textbf {\bibinfo {volume} {83}},\ \bibinfo {pages}
  {1002--1004} (\bibinfo {year} {2003})}\BibitemShut {NoStop}%
\bibitem [{\citenamefont {Gasparinetti}\ \emph {et~al.}(2015)\citenamefont
  {Gasparinetti}, \citenamefont {Viisanen}, \citenamefont {Saira},
  \citenamefont {Faivre}, \citenamefont {Arzeo}, \citenamefont {Meschke},\ and\
  \citenamefont {Pekola}}]{gasparinetti2015}%
  \BibitemOpen
  \bibfield  {author} {\bibinfo {author} {\bibfnamefont {S.}~\bibnamefont
  {Gasparinetti}}, \bibinfo {author} {\bibfnamefont {K.~L.}\ \bibnamefont
  {Viisanen}}, \bibinfo {author} {\bibfnamefont {O.-P.}\ \bibnamefont {Saira}},
  \bibinfo {author} {\bibfnamefont {T.}~\bibnamefont {Faivre}}, \bibinfo
  {author} {\bibfnamefont {M.}~\bibnamefont {Arzeo}}, \bibinfo {author}
  {\bibfnamefont {M.}~\bibnamefont {Meschke}}, \ and\ \bibinfo {author}
  {\bibfnamefont {J.~P.}\ \bibnamefont {Pekola}},\ }\bibfield  {title}
  {\enquote {\bibinfo {title} {Fast electron thermometry for ultrasensitive
  calorimetric detection},}\ }\href@noop {} {\bibfield  {journal} {\bibinfo
  {journal} {Phys. Rev. Appl.}\ }\textbf {\bibinfo {volume} {3}},\ \bibinfo
  {pages} {014007} (\bibinfo {year} {2015})}\BibitemShut {NoStop}%
\bibitem [{\citenamefont {Karimi}\ and\ \citenamefont
  {Pekola}(2018)}]{karimi2018}%
  \BibitemOpen
  \bibfield  {author} {\bibinfo {author} {\bibfnamefont {B.}~\bibnamefont
  {Karimi}}\ and\ \bibinfo {author} {\bibfnamefont {J.~P}\ \bibnamefont
  {Pekola}},\ }\bibfield  {title} {\enquote {\bibinfo {title} {Noninvasive
  thermometer based on the zero-bias anomaly of a superconducting junction for
  ultrasensitive calorimetry},}\ }\href@noop {} {\bibfield  {journal} {\bibinfo
   {journal} {Phys. Rev. Appl.}\ }\textbf {\bibinfo {volume} {10}},\ \bibinfo
  {pages} {054048} (\bibinfo {year} {2018})}\BibitemShut {NoStop}%
\bibitem [{\citenamefont {Karimi}\ \emph
  {et~al.}(2020{\natexlab{a}})\citenamefont {Karimi}, \citenamefont
  {Nikoli{\'c}}, \citenamefont {Tuukkanen}, \citenamefont {Peltonen},
  \citenamefont {Belzig},\ and\ \citenamefont {Pekola}}]{karimi2020pra}%
  \BibitemOpen
  \bibfield  {author} {\bibinfo {author} {\bibfnamefont {B.}~\bibnamefont
  {Karimi}}, \bibinfo {author} {\bibfnamefont {D.}~\bibnamefont {Nikoli{\'c}}},
  \bibinfo {author} {\bibfnamefont {T.}~\bibnamefont {Tuukkanen}}, \bibinfo
  {author} {\bibfnamefont {J.~T.}\ \bibnamefont {Peltonen}}, \bibinfo {author}
  {\bibfnamefont {W.}~\bibnamefont {Belzig}}, \ and\ \bibinfo {author}
  {\bibfnamefont {J.~P}\ \bibnamefont {Pekola}},\ }\bibfield  {title} {\enquote
  {\bibinfo {title} {Optimized proximity thermometer for ultrasensitive
  detection},}\ }\href@noop {} {\bibfield  {journal} {\bibinfo  {journal}
  {Phys. Rev. Appl.}\ }\textbf {\bibinfo {volume} {13}},\ \bibinfo {pages}
  {054001} (\bibinfo {year} {2020}{\natexlab{a}})}\BibitemShut {NoStop}%
\bibitem [{\citenamefont {Karimi}\ \emph
  {et~al.}(2020{\natexlab{b}})\citenamefont {Karimi}, \citenamefont {Brange},
  \citenamefont {Samuelsson},\ and\ \citenamefont
  {Pekola}}]{karimi2020natcomm}%
  \BibitemOpen
  \bibfield  {author} {\bibinfo {author} {\bibfnamefont {B.}~\bibnamefont
  {Karimi}}, \bibinfo {author} {\bibfnamefont {F.}~\bibnamefont {Brange}},
  \bibinfo {author} {\bibfnamefont {P.}~\bibnamefont {Samuelsson}}, \ and\
  \bibinfo {author} {\bibfnamefont {J.~P}\ \bibnamefont {Pekola}},\ }\bibfield
  {title} {\enquote {\bibinfo {title} {Reaching the ultimate energy resolution
  of a quantum detector},}\ }\href@noop {} {\bibfield  {journal} {\bibinfo
  {journal} {Nat. Commun.}\ }\textbf {\bibinfo {volume} {11}},\ \bibinfo
  {pages} {1--6} (\bibinfo {year} {2020}{\natexlab{b}})}\BibitemShut {NoStop}%
\bibitem [{\citenamefont {Zhou}\ \emph {et~al.}(1998)\citenamefont {Zhou},
  \citenamefont {Charlat}, \citenamefont {Spivak},\ and\ \citenamefont
  {Pannetier}}]{zhou1998}%
  \BibitemOpen
  \bibfield  {author} {\bibinfo {author} {\bibfnamefont {F.}~\bibnamefont
  {Zhou}}, \bibinfo {author} {\bibfnamefont {P.}~\bibnamefont {Charlat}},
  \bibinfo {author} {\bibfnamefont {B.}~\bibnamefont {Spivak}}, \ and\ \bibinfo
  {author} {\bibfnamefont {B.}~\bibnamefont {Pannetier}},\ }\bibfield  {title}
  {\enquote {\bibinfo {title} {Density of states in superconductor-normal
  metal-superconductor junctions},}\ }\href@noop {} {\bibfield  {journal}
  {\bibinfo  {journal} {J. Low Temp. Phys.}\ }\textbf {\bibinfo {volume}
  {110}},\ \bibinfo {pages} {841--850} (\bibinfo {year} {1998})}\BibitemShut
  {NoStop}%
\bibitem [{\citenamefont {Le~Sueur}\ \emph {et~al.}(2008)\citenamefont
  {Le~Sueur}, \citenamefont {Joyez}, \citenamefont {Pothier}, \citenamefont
  {Urbina},\ and\ \citenamefont {Est\`eve}}]{lesueur2008}%
  \BibitemOpen
  \bibfield  {author} {\bibinfo {author} {\bibfnamefont {H.}~\bibnamefont
  {Le~Sueur}}, \bibinfo {author} {\bibfnamefont {P.}~\bibnamefont {Joyez}},
  \bibinfo {author} {\bibfnamefont {H.}~\bibnamefont {Pothier}}, \bibinfo
  {author} {\bibfnamefont {C.}~\bibnamefont {Urbina}}, \ and\ \bibinfo {author}
  {\bibfnamefont {D.}~\bibnamefont {Est\`eve}},\ }\bibfield  {title} {\enquote
  {\bibinfo {title} {Phase controlled superconducting proximity effect probed
  by tunneling spectroscopy},}\ }\href@noop {} {\bibfield  {journal} {\bibinfo
  {journal} {Phys. Rev. Lett.}\ }\textbf {\bibinfo {volume} {100}},\ \bibinfo
  {pages} {197002} (\bibinfo {year} {2008})}\BibitemShut {NoStop}%
\bibitem [{\citenamefont {Ligato}\ \emph {et~al.}(2021)\citenamefont {Ligato},
  \citenamefont {Strambini}, \citenamefont {Paolucci},\ and\ \citenamefont
  {Giazotto}}]{ligato2021}%
  \BibitemOpen
  \bibfield  {author} {\bibinfo {author} {\bibfnamefont {N.}~\bibnamefont
  {Ligato}}, \bibinfo {author} {\bibfnamefont {E.}~\bibnamefont {Strambini}},
  \bibinfo {author} {\bibfnamefont {F.}~\bibnamefont {Paolucci}}, \ and\
  \bibinfo {author} {\bibfnamefont {F.}~\bibnamefont {Giazotto}},\ }\bibfield
  {title} {\enquote {\bibinfo {title} {Preliminary demonstration of a
  persistent {J}osephson phase-slip memory cell with topological protection},}\
  }\href@noop {} {\bibfield  {journal} {\bibinfo  {journal} {Nat. Commun.}\
  }\textbf {\bibinfo {volume} {12}},\ \bibinfo {pages} {1--8} (\bibinfo {year}
  {2021})}\BibitemShut {NoStop}%
\bibitem [{\citenamefont {Larkin}\ and\ \citenamefont
  {Ovchinnikov}(1968)}]{LarkinOvchinnikov1968}%
  \BibitemOpen
  \bibfield  {author} {\bibinfo {author} {\bibfnamefont {A.~L.}\ \bibnamefont
  {Larkin}}\ and\ \bibinfo {author} {\bibfnamefont {Yu.~N.}\ \bibnamefont
  {Ovchinnikov}},\ }\bibfield  {title} {\enquote {\bibinfo {title}
  {Quasiclassical method in the theory of superconductivity},}\ }\href@noop {}
  {\bibfield  {journal} {\bibinfo  {journal} {Zh. Eksp. Teor. Fiz.}\ }\textbf
  {\bibinfo {volume} {55}},\ \bibinfo {pages} {2262} (\bibinfo {year}
  {1968})}\BibitemShut {NoStop}%
\bibitem [{\citenamefont {Usadel}(1970)}]{Usadel1970}%
  \BibitemOpen
  \bibfield  {author} {\bibinfo {author} {\bibfnamefont {K.~D.}\ \bibnamefont
  {Usadel}},\ }\bibfield  {title} {\enquote {\bibinfo {title} {Generalized
  diffusion equation for superconducting alloys},}\ }\href@noop {} {\bibfield
  {journal} {\bibinfo  {journal} {Phys. Rev. Lett.}\ }\textbf {\bibinfo
  {volume} {25}},\ \bibinfo {pages} {507} (\bibinfo {year} {1970})}\BibitemShut
  {NoStop}%
\bibitem [{\citenamefont {Belzig}\ \emph {et~al.}(1999)\citenamefont {Belzig},
  \citenamefont {Wilhelm}, \citenamefont {Bruder}, \citenamefont {Sch\"on},\
  and\ \citenamefont {Zaikin}}]{Belzig1999}%
  \BibitemOpen
  \bibfield  {author} {\bibinfo {author} {\bibfnamefont {W.}~\bibnamefont
  {Belzig}}, \bibinfo {author} {\bibfnamefont {F.~K.}\ \bibnamefont {Wilhelm}},
  \bibinfo {author} {\bibfnamefont {C.}~\bibnamefont {Bruder}}, \bibinfo
  {author} {\bibfnamefont {G.}~\bibnamefont {Sch\"on}}, \ and\ \bibinfo
  {author} {\bibfnamefont {A.~D.}\ \bibnamefont {Zaikin}},\ }\bibfield  {title}
  {\enquote {\bibinfo {title} {Quasiclassical {G}reen's function approach to
  mesoscopic superconductivity},}\ }\href@noop {} {\bibfield  {journal}
  {\bibinfo  {journal} {Superlattices Microstruct.}\ }\textbf {\bibinfo
  {volume} {25}},\ \bibinfo {pages} {1251} (\bibinfo {year}
  {1999})}\BibitemShut {NoStop}%
\bibitem [{\citenamefont {Giazotto}\ \emph {et~al.}(2006)\citenamefont
  {Giazotto}, \citenamefont {Heikkil{\"a}}, \citenamefont {Luukanen},
  \citenamefont {Savin},\ and\ \citenamefont
  {Pekola}}]{giazotto2006opportunities}%
  \BibitemOpen
  \bibfield  {author} {\bibinfo {author} {\bibfnamefont {F.}~\bibnamefont
  {Giazotto}}, \bibinfo {author} {\bibfnamefont {T.~T.}\ \bibnamefont
  {Heikkil{\"a}}}, \bibinfo {author} {\bibfnamefont {A.}~\bibnamefont
  {Luukanen}}, \bibinfo {author} {\bibfnamefont {A.~M.}\ \bibnamefont {Savin}},
  \ and\ \bibinfo {author} {\bibfnamefont {J.~P.}\ \bibnamefont {Pekola}},\
  }\bibfield  {title} {\enquote {\bibinfo {title} {Opportunities for
  mesoscopics in thermometry and refrigeration: Physics and applications},}\
  }\href@noop {} {\bibfield  {journal} {\bibinfo  {journal} {Rev. Mod. Phys.}\
  }\textbf {\bibinfo {volume} {78}},\ \bibinfo {pages} {217} (\bibinfo {year}
  {2006})}\BibitemShut {NoStop}%
\bibitem [{\citenamefont {Viisanen}\ and\ \citenamefont
  {Pekola}(2018)}]{viisanen2018}%
  \BibitemOpen
  \bibfield  {author} {\bibinfo {author} {\bibfnamefont {K.~L.}\ \bibnamefont
  {Viisanen}}\ and\ \bibinfo {author} {\bibfnamefont {J.~P.}\ \bibnamefont
  {Pekola}},\ }\bibfield  {title} {\enquote {\bibinfo {title} {Anomalous
  electronic heat capacity of copper nanowires at sub-kelvin temperatures},}\
  }\href@noop {} {\bibfield  {journal} {\bibinfo  {journal} {Phys. Rev. B}\
  }\textbf {\bibinfo {volume} {97}},\ \bibinfo {pages} {115422} (\bibinfo
  {year} {2018})}\BibitemShut {NoStop}%
\bibitem [{\citenamefont {Wang}\ \emph {et~al.}(2019)\citenamefont {Wang},
  \citenamefont {Golubev}, \citenamefont {Galperin},\ and\ \citenamefont
  {Pekola}}]{wang2019}%
  \BibitemOpen
  \bibfield  {author} {\bibinfo {author} {\bibfnamefont {L.~B.}\ \bibnamefont
  {Wang}}, \bibinfo {author} {\bibfnamefont {D.~S.}\ \bibnamefont {Golubev}},
  \bibinfo {author} {\bibfnamefont {Y.~M.}\ \bibnamefont {Galperin}}, \ and\
  \bibinfo {author} {\bibfnamefont {J.~P.}\ \bibnamefont {Pekola}},\ }\bibfield
   {title} {\enquote {\bibinfo {title} {Dynamic thermal relaxation in metallic
  films at sub-kelvin temperatures},}\ }\href@noop {} {\bibfield  {journal}
  {\bibinfo  {journal} {arXiv preprint arXiv:1910.09448}\ } (\bibinfo {year}
  {2019})}\BibitemShut {NoStop}%
\bibitem [{\citenamefont {Pekola}\ and\ \citenamefont
  {Karimi}(2022)}]{pekola2022}%
  \BibitemOpen
  \bibfield  {author} {\bibinfo {author} {\bibfnamefont {J.~P.}\ \bibnamefont
  {Pekola}}\ and\ \bibinfo {author} {\bibfnamefont {B.}~\bibnamefont
  {Karimi}},\ }\bibfield  {title} {\enquote {\bibinfo {title} {Ultrasensitive
  calorimetric detection of single photons from qubit decay},}\ }\href@noop {}
  {\bibfield  {journal} {\bibinfo  {journal} {Phys. Rev. X}\ }\textbf {\bibinfo
  {volume} {12}},\ \bibinfo {pages} {011026} (\bibinfo {year}
  {2022})}\BibitemShut {NoStop}%
\bibitem [{\citenamefont {Braine}\ \emph {et~al.}(2020)\citenamefont {Braine},
  \citenamefont {Cervantes}, \citenamefont {Crisosto}, \citenamefont {Du},
  \citenamefont {Kimes}, \citenamefont {Rosenberg}, \citenamefont {Rybka},
  \citenamefont {Yang}, \citenamefont {Bowring}, \citenamefont {Chou} \emph
  {et~al.}}]{braine2020}%
  \BibitemOpen
  \bibfield  {author} {\bibinfo {author} {\bibfnamefont {T.}~\bibnamefont
  {Braine}}, \bibinfo {author} {\bibfnamefont {R.}~\bibnamefont {Cervantes}},
  \bibinfo {author} {\bibfnamefont {N.}~\bibnamefont {Crisosto}}, \bibinfo
  {author} {\bibfnamefont {N.}~\bibnamefont {Du}}, \bibinfo {author}
  {\bibfnamefont {S.}~\bibnamefont {Kimes}}, \bibinfo {author} {\bibfnamefont
  {L.~J.}\ \bibnamefont {Rosenberg}}, \bibinfo {author} {\bibfnamefont
  {G.}~\bibnamefont {Rybka}}, \bibinfo {author} {\bibfnamefont
  {J.}~\bibnamefont {Yang}}, \bibinfo {author} {\bibfnamefont {D.}~\bibnamefont
  {Bowring}}, \bibinfo {author} {\bibfnamefont {A.~S.}\ \bibnamefont {Chou}},
  \emph {et~al.},\ }\bibfield  {title} {\enquote {\bibinfo {title} {Extended
  search for the invisible axion with the axion dark matter experiment},}\
  }\href@noop {} {\bibfield  {journal} {\bibinfo  {journal} {Phys. Rev. Lett.}\
  }\textbf {\bibinfo {volume} {124}},\ \bibinfo {pages} {101303} (\bibinfo
  {year} {2020})}\BibitemShut {NoStop}%
\end{thebibliography}%
%\end{thebibliography}
\end{document}